# Precompensation of 3D field distortions in remote focus two-photon microscopy

Antoine M. Valera[1,†], Fiona C. Neufeldt[1,2†], Paul A. Kirkby[1], John E. Mitchell[2] and R. Angus Silver[1,*]

[1]*Department of Neuroscience, Physiology and Pharmacology, University College London, Gower Street, London WC1E 6BT, UK*
[2]*Department of Electronic and Electrical Engineering, University College London, Malet Place, London WC1E 7JE, UK*
*\*Corresponding address: a.silver@ucl.ac.uk*

† *These authors contributed equally.*

**Remote focusing is widely used in 3D two-photon microscopy and 3D photostimulation because it enables fast axial scanning without moving the objective lens or specimen. However, due to the design constraints of microscope optics, remote focus units are often located in non-telecentric positions in the optical path, leading to significant depth-dependent 3D field distortions in the imaging volume. To address this limitation, we characterized 3D field distortions arising from non-telecentric remote focusing and present a method for distortion precompensation. We demonstrate its applicability for a 3D two-photon microscope that uses an acousto-optic lens (AOL) for remote focusing and scanning. We show that the distortion precompensation method improves the pointing precision of the AOL microscope to < 0.5 μm throughout the 400 ×400 × 400 μm imaging volume.**

## 1. INTRODUCTION

Two-photon scanning microscopy is a widely used tool for high resolution functional imaging in life sciences due to its ability to penetrate deep within scattering tissue [1,2]. In neuroscience, two-photon imaging is often used to monitor activity within neurons and across neuronal populations. However, measurements of signals flowing through these 3D structures are restricted with conventional microscopes because the speed of focusing is limited by the inertia of the objective lens. By contrast, rapid remote focus approaches rely on devices that dynamically alter the optical wavefront of the beam prior to the objective, thereby circumventing the need to physically move the objective. Rapid remote focusing has been implemented with electrically tunable lenses (ETLs) [3], liquid crystal spatial light modulators (LC-SLMs) [4–8], phase-locked ultrasound lenses (tunable acoustic gradient (TAG) lenses) [9], deformable mirrors [10], acousto-optic lenses (AOLs) [11–14] and a secondary objective coupled with a piston mirror [15–19]. However, remote focus devices cannot always be positioned in a plane conjugate to the back aperture of the objective, due to mechanical space constraints, a limited range of commercially available focal length lenses and uncertainty in the exact position of the back aperture of commercial objectives. This is problematic because it introduces depth-dependent 3D field distortions in the imaging volume [3,20–23], which impair the performance of remote focus laser scanning systems. Distortions to the 3D field of view (FOV) lead to missed regions of the specimen and serious mismatches between the 3D *z*-stacks obtained by mechanically moving the objective and the *z*-stacks obtained by remote focusing. It also causes erroneous measurements of the dimensions of biological features [20] and makes selection and positioning of regions of interest (ROIs) for selective imaging or photostimulation, within the full FOV inaccurate, which is particularly detrimental when targeting small structures such as neuronal processes.

In a conventional microscope with a collimated input and an infinity corrected objective, non-telecentricity has little effect on the 2D FOV when the focus is changed by mechanically shifting the objective. Although a lateral misalignment of the input beam will result in a tilted point spread function (PSF), changing the focus will not introduce *xy* plane magnification distortions. However, in a remote focus microscope, where the curvature and tilt of the input beam is modified to change the 3D focal position, a misalignment of the optical input from its ideal telecentric position leads to a distorted 3D FOV. An axially misaligned optical input leads to a *z*-dependent *xy* magnification, as well as a varying axial magnification, resulting in uneven *z*-plane separation. A lateral misalignment of the remote focus unit results in an axially skewed (*xz* and/or *yz*) FOV. As the laser is focused away in *z* from the natural plane (the focal plane for a planar input beam), the FOV progressively drifts laterally.

Previous distortion correction approaches [20] have been implemented via post-acquisition image processing. This has several limitations, including loss of information due to cropping of the volumetric data, the time-consuming nature of *post-hoc* corrections and the use of a complex calibration sample which requires laser manufacturing and which is sensitive to misalignments. Here, we present a fast, convenient, and generic method to measure and precompensate distortions in remote focus microscopes. We demonstrate experimentally that a paraxial remote focus distortion model can be applied to an AOL microscope to precompensate for the 3D field distortions arising from non-telecentric misalignments and show how similar distortion precompensation solutions could be applied to other remote focus technologies.

## 2. MODEL OF 3D FIELD DISTORTIONS IN REMOTE FOCUS MICROSCOPES

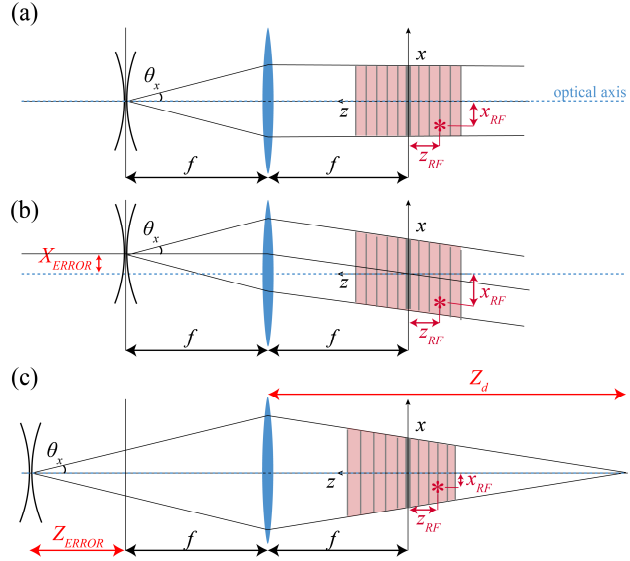

**Fig. 1. Telecentricity in remote focus microscopes**. **a)** Schematic ray diagram showing *xz* plane for telecentric output of the wavefront shaping unit located one focal length (*f*) from the imaging objective (*blue* lens). Curved wavefronts (*curved black* lines) represent a converging or diverging optical beam. Central ray (*black* lines) deflected through a beam semi-scan angle $\theta_x$, with respect to the optical axis (*dotted blue* line). Each of the three central rays indicates a different time point of a scan. Arbitrary *xz* remote focus (*red* asterisks) within the FOV (*pink* area) with coordinates ($x_{RF}$, $z_{RF}$) indicated. The change in the axial focus of the objective is linearly proportional to the change in curvature produced by the remote focus unit and the central ray in the FOV has no skew. Imaging planes (*gray*) are thus equally spaced in *z* for equal steps in beam curvature and the FOV is of constant size. **b)** Lateral misalignment of the remote focus device output ($X_{ERROR}$) produces a skewed FOV, resulting in a skewed local *z*-axis and a shifted $x_{RF}$. **c)** Axial misalignment of the remote focus device output ($Z_{ERROR}$) causes depth-dependent *xy* FOV magnification and a non-constant *z* plane spacing with wavefront curvature, resulting in a change in both $x_{RF}$ and $z_{RF}$. $Z_d$ is the position of the plane conjugate with the output of the remote focus device.

To characterize and correct for the 3D field distortions that arise in a remote focus microscope due to non-telecentric misalignments of the remote focus unit, we developed a paraxial model of a remote focus unit and objective lens. In an ideal telecentric case [Fig. 1(a)] the position of the output of the remote focus unit is located one focal length distance away from the objective and the optical curvature change produced by the remote focus unit is linearly proportional to the resulting change of *z*-remote-focus ($z_{RF}$) in the objective FOV.

Note that $z_{RF}$ is measured with respect to the natural focal plane and is considered a negative distance when located below the natural focal plane (as is the case in Fig. 1). Since the plane conjugate to the output of the remote focus unit is located at infinity, the magnification of the *xy* FOV of the objective is also constant with changes in axial focus. A lateral displacement in *x*, $X_{ERROR}$, of the output of the remote focus unit introduces an *xz* skew to the objective FOV [Fig. 1(b)], shifting the position of the *x*-remote-focus ($x_{RF}$) with *z*. Similarly, a lateral offset in *y* would cause a *yz* skew. An axial misalignment of the remote focus unit, $Z_{ERROR}$, results in a *z*-dependent *xy* magnification of the objective FOV, together with a non-uniform *z*-plane separation for equal changes in curvature of the optical input [Fig. 1(c)]. In this case, the change in position of $z_{RF}$ is not proportional to the change in optical curvature produced by the remote focus device and the resulting *z*-dependent *xy* magnification causes tapering of the *x* FOV with *z*.

To quantify the field distortions arising from non-telecentric misalignments of the remote focus unit we derived a set of equations (see Supplement 1 Section S1 and [Fig. S1] for the full derivation). For an optical output of curvature $\kappa$, a paraxial objective lens of focal length *f*, an imaging refractive index *n* and an (*x*, *z*) misalignment of the remote focus unit ($X_{ERROR}$, $Z_{ERROR}$), the (*x*, *z*) remote focus with respect to the optical axis, ($x_{RF}$, $z_{RF}$) can be given by,

$$z_{RF} = \frac{-n f^2 \kappa}{\kappa Z_{ERROR} + 1} \qquad (1)$$

$$x_{RF} = \frac{f(\theta_x - \kappa X_{ERROR})}{\kappa Z_{ERROR} + 1} \qquad (2)$$

where $\theta_x$ is the uncompensated *x* semi-scan angle. Note that Fig. 1 shows the deflection of the central ray of the wavefront at the beginning, middle and end of a scan. For an axially misaligned remote focus unit, the lateral *xy* plane magnification *M*, will vary with *z* and is given by,

$$M = \frac{nf^2 + z_{RF} Z_{ERROR}}{nf^2} \qquad (3)$$

To precompensate for the distortion, the remote focus unit can be driven to produce a wavefront curvature $\kappa_{COMP}$ and $x$ semi-scan angle $\theta_{x\,COMP}$, based on the inverse of equations 1 and 2, for the desired corrected $x$ and $z$ focus in the objective FOV, $z_{CORR}$ and $x_{CORR}$, respectively, where

$$\kappa_{COMP} = - \frac{z_{CORR}}{z_{CORR} Z_{ERROR} + nf^2} \qquad (4)$$

$$\theta_{X\,COMP} = \kappa_{COMP} X_{ERROR} + \frac{x_{CORR} (\kappa_{COMP} Z_{ERROR} + 1)}{f} \qquad (5)$$

Thus, distortions arising from system non-telecentricity can be perfectly compensated, if the wavefront curvature $\kappa$ and $(x, y)$ semi scan angles $(\theta_x, \theta_y)$ produced by the remote focus unit are known. The generic nature of this solution suggests that field distortions caused by any paraxially approximated remote focus system can be precompensated by incorporating equations 4 and 5 into the control software. Magnification distortions are common in widely used ETL-based remote focus microscopes because they are often placed in a non-telecentric position behind the objective [21–23]. As a worked example, we show how our solution could be used to precompensate the remote focus distortion reported in an ETL-based non-telecentric remote focus two-photon microscope [3]. Fig 2(a) shows the reported FOV and 2(b) the FOV fitted by eqs. (1,2). The precompensated FOV calculated by eqs. (4,5) is shown in Fig. 2(c). By driving the non-telecentric remote focus unit using eqs. (4,5), a perfectly undistorted FOV would be obtained, as predicted in Fig. 2(d).

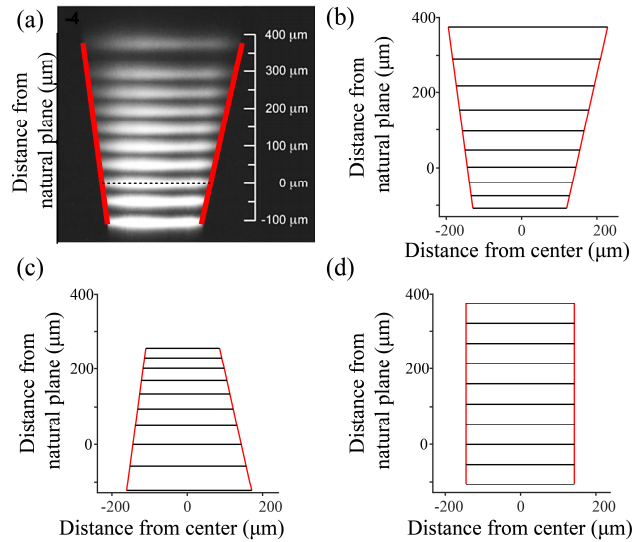

**Fig. 2. Application of the distortion model to an electrically tunable lens remote focus system. a)** Experimental $xz$ FOV imaged in fluorescein beneath the objective of a two-photon microscope, remote focused with an electrically tunable lens. Adapted with permission from [3] © The Optical Society. *Red* lines show extent of $x$ limits of the FOV predicted with fitted using eqs. (1,2) with $f = 4$ mm, $Z_{ERROR} = 26.6$ mm, $X_{ERROR} = 0.4$ mm, $n = 1.33$ and $\theta_x = \pm 36$ mrad. Natural focal plane ($z = 0$) indicated by dashed black line. **b)** Theoretical $xz$ FOV for a $z$ range = [-108, 375] μm with 9 equal steps of curvature, given eqs. (1,2). **c)** Theoretical $xz$ FOV calculated using precompensated drives $\kappa_{COMP}$, eq. (4) and $\theta_{X\,COMP}$. **d)** Accurately precompensated rectangular FOV with equispaced $z$ planes calculated by driving the non-telecentric model of the remote focus system of a) and b) with the pre-compensated drive of c).

## 3. EXPERIMENTAL EVALUATION OF THE DISTORTION PRECOMPENSATION SCHEME

To test the accuracy of distortion precompensation we developed a simple strategy [Fig. 3(a)] to measure field distortions and employed an acousto-optic lens (AOL) remote focus microscope [Fig. S2] (Supplement 1 Section S2), since this technology enables high precision focusing and scanning of a laser beam [24,25]. To quantify distortions caused by non-telecentric remote focusing in our AOL 3D two-photon microscope we compared a reference image obtained at the natural plane of the objective [Fig. 3(b)], where non-telecentric misalignment does not cause distortion of the FOV, with images of the same region obtained using increasingly strong remote focusing. The resulting change in focus was then counteracted by displacing the objective by an equal amount in the opposite direction [Fig. 3(a), inset]. The precision of the mechanical objective focus, which is critical to the precision of the calibration scheme, was determined to be <0.5 μm, with a mechanical alignment gauge. To perform the calibration, we used a microscope slide with a thin layer of 5 μm fluorescent beads suspended in agar (Phosphorex, Degradex), although any small, bright sample would work (see Supplement 1 Section S3 and [Fig. S3]). In a perfectly telecentric optical path, the mechanical objective focus and AOL remote focus would be in perfect registration and the images of the beads would be superimposed wherever they were acquired across the entire remote focusing *z*-range. However, when non-telecentric misalignment is present, field distortions are revealed as displacements in the locations of the beads in the images across the *z*-range. Fig. 3(c) shows a *z*-intensity projection of the images of the calibration *z*-stack (*C-z*-stack) covering a remote focus range of ± 200 µm, without correction for the distortion arising from the misalignments present in our AOL 3D two-photon microscope. Displacement of the individual beads across the *C-z*-stack were measured using the TrackMate plugin in ImageJ [26]. When plotting the position of the beads for each frame of the *C-z*-stack, the trajectory of the displacement of each bead over the remote focus *z*-range enables a measure of the distortion field [Fig. 3(d)].

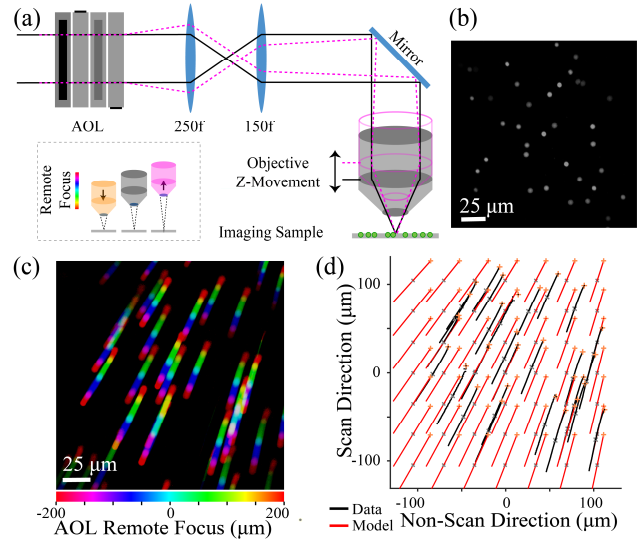

**Fig. 3. Characterization of non-telecentric remote focus distortions in an acousto-optic lens microscope**. a) Experimental setup, showing a slide covered with a thin layer of 5 µm green beads. As the axial position of the objective is incrementally shifted away from the sample using motorized *z*-movement, the beads are kept in focus using remote focusing with an acousto-optic lens (AOL) microscope. *Black solid line* shows optical beam focusing at the natural plane (*gray objective*). *Pink dotted line shows* converging optical beam enabling remote focused imaging with objective displaced (*pink outlined objective*). *Inset* summarizes the experimental procedure used to generate panel (c). **b)** Reference image of 5 µm fluorescent beads obtained at the natural focus of the objective in a 262 µm *xy* FOV. **c)** Color-coded intensity projection of the imaged beads at various remote foci (± 200 µm in 5 µm increments, color indicates *z*-plane location). Color-coded lines indicate depth-dependent deviation of the *xy* bead location relative to the natural plane image. **d)** Bead trajectories from (c) in scanned (*y*) and non-scanned (*x*) dimensions tracked using ImageJ TrackMate plugin (black). Bead trajectories predicted from the model (red). *Orange and black asterisks* indicate location of beads for +200 µm remote focusing and natural plane imaging, respectively.

To relate bead trajectories to non-telecentric field distortion in a quantitative manner we developed a ray-based model of the AOL 3D microscope. The non-telecentric remote focus distortion predicted by the AOL ray model was found to match the distortion predicted by the paraxial distortion model [Fig. S4-S6]. The ray model was then used to estimate the values of non-telecentric misalignment in the physical system that give rise to the observed distortion, by finding a match between the experimentally measured distortion and the distortion predicted by the model. For the experimental data in Fig. 3(d), the distortion predicted by the model is likely to be caused by a misalignment of the AOL scanner of $X_{ERROR}$, $Y_{ERROR}$, $Z_{ERROR}$ of 1.4, 0.4, 69 mm, respectively [Fig. S7]. But note that the same experimental field distortion could be caused by a wide range of optical component misalignments in the relay(s) between the remote focus device and the final objective. Nevertheless, the predictions of eqs. 1 and 2 will still be valid. We then used (eqs. 4,5) to precompensate for the distortion, by modifying the amount of tilt and curvature imparted by the AOL to the optical beam. The distortion precompensation equations were incorporated into our MATLAB-based AOL microscope control software and had little computational overhead. The values for $X_{ERROR}$, $Y_{ERROR}$ and $Z_{ERROR}$ were adjusted to obtain a perfect registration between the remote focus system and the mechanical focus system.
The performance of the distortion precompensation scheme was tested for a range of effective lateral and axial misalignments of the AOL by using a pair of Risley prisms (PS810-B, Thorlabs Inc.) to mimic the distortion produced by lateral misalignments of the AOL scanner.
The prisms were placed in the path between the relay lenses, to laterally offset the beam in a controlled manner [Fig. S8]. The effective lateral misalignment of the AOL, $X_{ERROR}$ and $Y_{ERROR}$ could be calculated from the lateral displacement of the optical beam by the Risley prisms and the magnification of the optical telecentric relay. To assess the performance of the distortion correction scheme, *xz* and *yz* skew of the FOV was measured by taking the inverse tangent of the slope of the trajectory of the beads through the *C-z*-stacks, with and without the distortion correction scheme enabled, for a range of lateral beam displacements. Fig. 4(a,b) shows the maximum intensity *z*-projections of the *C-z*-stacks acquired without and with distortion precompensation, for an effective AOL misalignment of 6.2 mm in *y*. Despite the strong distortion introduced by the *y* misalignment (-19.3 ± 0.1 degree

*yz* skew angle over a ±200 µm remote focus range), the distortion precompensation was capable of reducing the *yz* skew angle to -0.0±0.1 degrees. Risley prism-based lateral beam displacement across a range of values showed that the distortion compensation scheme was effective at correcting skew distortions arising from effective displacements of the remote focus unit of up to ±7 mm in *y* [Fig 4(c)]. Similar results were obtained for *x* displacement. To test the effectiveness of precompensation for magnification distortions arising from axial misalignments of the AOL scanner, the amount of effective axial misalignment of the AOL scanner, $Z_{ERROR}$ was varied by changing the lenses within the relay, which altered the magnification and introduced an effective axial misalignment of 152 mm and 9 mm (i.e. an addition of 83 mm and -60 mm to the pre-existing axial misalignment of 69 mm in our AOL setup) [Fig. S9]. Fig. S10 shows the correction of these effective axial misalignments, confirming the capacity of the distortion precompensation scheme for larger axial misalignments of the remote focus unit. In addition, the *z* focal plane position was measured before and after distortion precompensation, confirming the correction of the non-constant *z*-spacing introduced by the pre-existing axial misalignment of 69 mm [Fig. S11]. To quantify the precision of AOL-based laser scanning throughout the 400 × 400 × 400 µm imaging volume after distortion precompensation, we measured the largest displacement of the bead from its position when imaged at the natural focal plane of the objective (*z* = 0). Fig. 4(d) provides an example of the bead displacement measured over a *C-z*-stack, *after* correction for the inherent lateral and axial misalignments of the AOL. This revealed a maximum lateral displacement of 0.33 µm ± 0.12 µm over the ±200 µm remote focus range.

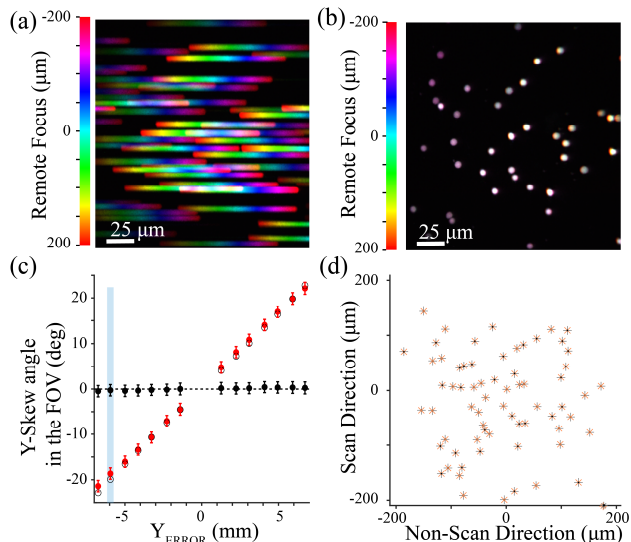

**Fig. 4. Distortion precompensation: a)** Color-coded maximal intensity projection of fluorescent beads for a ±200 µm remote focusing range with 5 µm *z*-increments, without correction and with the beam incident on the back aperture of the objective misaligned using the Risley prisms (equivalent to a -6.2 mm $Y_{ERROR}$). **b)** Same as (a) with distortion precompensation. **c)** YZ skew angle in the FOV of the objective (arctan of the *yz* slope of the central bead) as a function of the effective lateral AOL misalignment in *y* (*red symbols*). The distortion due to the inherent axial misalignment of the AOL was precompensated using the correction scheme to isolate the distortion arising from the Risley prism beam deflection (*black* symbols). Theoretical skew measurements (*hollow black* circles). *Light blue* area indicates the experimental point shown in (b, c). **d)** Trajectories of the bead positions (*black*) with both axial and lateral precompensation. *Orange* and *gray* asterisks indicate location of beads values for +200 µm remote focusing and natural plane imaging, respectively. Note that markers and lines are superimposed as residual error is only 0.3 ± 0.2 µm.

An unexpected benefit of the high precision of the distortion precompensation scheme for non-telecentric misalignments enabled us to identify and correct for small residual distortions in our AOL microscope that arose from weak divergence of the input beam [Fig. S12]. This higher order distortion, which resulted in a non-uniform distortion of the field with a scan-time dependency, could be explained by the ray model (Supplement 1 Section S6), enabling a full correction. This was possible only after precompensation for the larger first order distortions due to system non-telecentricity.

## 4. DISCUSSION AND CONCLUSION
Wavefront-shaping devices are increasingly being used in two-photon microscopy to provide the rapid focusing required for 3D functional imaging and 3D photostimulation. Here we derive a set of equations to quantify and correct for the substantial field distortions introduced by remote focus devices that are not perfectly telecentric. By quantifying distortions and implementing distortion precompensation into the control software of an AOL-based remote focus 3D two-photon microscope, we demonstrated submicron (< 0.5 µm) precision over the 400 × 400 × 400 µm FOV, for a non-telecentric arrangement. The generic nature of the precompensation solution makes it suitable for a range of remote focus devices, enabling non-telecentric designs with improved optical performance.

Addition of remote focusing devices to conventional microscopes is complicated by the requirement for telecentricity within the optical train [27], which is often not possible due to mechanical space constraints, lens availability and a lack of information on the optical design of commercial microscopes and objectives. The substantial field distortions introduced by non-telecentric arrangements [3,20–23] compromise the imaging performance of such 3D microscopes. Our solution to this problem, which enables full precompensation of distortion over the entire 3D FOV, builds on earlier work that quantified remote focus distortions using a sophisticated 3D plastic reference calibration specimen and an image processing based *post-hoc* correction [20]. Our quantification method and precompensation strategy has several advantages over such an approach. The simple *C-z*-stack calibration method using fluorescent beads dispenses with the need for a custom made 3D calibration sample, and thus any issues associated with the polymer having a different refractive index to tissue and hence a different distortion (as noted by the authors, [20]). By contrast our calibration and precompensation can be carried out in water or even in the target biological tissue itself. This allows any optical element located between the lens and the sample, such as a coverslip, to be integrated into the calibration process. Moreover, our approach directly calibrates the microscope against the *xyz* stage mechanics and thus provides a single coordinate system for the 3D FOV of the specimen when zooming and/or imaging in any arbitrary plane. Lastly, our precompensation solution restores the full telecentric imaging volume, so that remote focusing and mechanical focusing are in perfect register and laser scanning in any arbitrary direction maps accurately onto a Euclidean 3D space. This prevents regions being missed and eliminates the necessity of post processing the 3D images.

Quantification and precompensation of field distortions, represents a significant step towards achieving aberration-free non-telecentric remote focusing. Compensation for field distortions does not, however, correct for the (largely spherical) aberrations introduced by the objective lens when remote focusing. These have recently been characterized in detail and compensated for in a LC-SLM-remote focus microscope [28]. Combining field and lens-based aberration corrections would enable non-telecentric remote focus designs to approach the near optimal optical performance achieved with telecentric dual objective 3D microscopes [15,19].

Precompensation of remote focus field distortion improves the performance of AOL-based random-access 3D microscopy in several ways. Firstly, it ensures that the location and extent of the FOV is the same as with a perfectly telecentric optical arrangement. The sub 0.5 µm precision achievable with AOL-based laser scanning is critical for selective high-speed functional imaging of fine biological structures such as the synapses, axons and dendrites of neurons. High precision line scanning within an undistorted imaging volume is also critical for the correction of tissue movement in real time, since this involves tracking the movement of an object such as a fluorescent bead, in one location and applying a rigid *xyz* translation to the imaged volume, to counteract any 3D brain movement [29]. Lastly, an undistorted remote focus FOV enables highly precise alignment of neighboring AOL-based *z*-stacks for imaging large biological structures and brain regions that extend far beyond a single imaging volume.

As our modelling of ETL remote focusing shows, our approach for correcting distortions in non-telecentric remote focus devices is applicable to other types of 3D microscopes and is expected to improve their performance. Precompensation for field distortions is also likely to be particularly important for 3D photostimulation, since this requires accurate focusing to one or more precise locations within the imaging volume. Indeed, 3D two-photon imaging and optogenetic photostimulation methods [7,30–33] require accurate LC-SLM-remote focusing of multiple illumination beams to specific locations within the imaging volume, which is typically acquired with galvanometer-based 2D scanning and mechanical focusing. Precise precompensation of the imaging field will be critical for extending these technologies to combine full 3D random access functional imaging and 3D photostimulation and for making photostimulation more precise, so that smaller structures such as synapses (~1µm) can be selectively photoactivated.

**FUNDING SOURCES, ACKNOWLEDGMENTS, AND DISCLOSURES**

Supported by the Wellcome Trust (203048; R.A.S. is in receipt of a Wellcome Trust Principal Research Fellowship in Basic Biomedical Science) and the National Institute of Neurological Disorders and Stroke of the National Institutes of Health under award number U01NS113273 (to R.A.S.). The content is solely the responsibility of the authors and does not necessarily represent the official views of the National Institutes of Health. FN was supported by the UK Engineering and Physical Sciences Research Council (EPSRC) studentship. We thank Victoria Griffiths, K. M. Naga Srinivas Nadella, Tomás Fernández-Alfonso for comments on the manuscript. A.M.V., F.C.N., J.E.M declare no competing interests. P.A.K. and R.A.S. are named inventors on patents owned by UCL Business relating to linear and nonlinear AOL 3D laser scanning technology. P.A.K. has a financial interest in Agile Diffraction Ltd., which aims to commercialize the AOL technology.

This document contains supplementary information to "Precompensation of 3D field distortions in remote focus two-photon microscopy". It provides the derivation of the equations describing the non-telecentric distortion model (section S1) and includes details on the optical path of an acousto-optic lens (AOL) remote focus two-photon microscope (section S2). An outline of the method of implementing distortion precompensation with a single fluorescent bead is given in section S3. Section S4 gives examples of the 3D field distortions predicted by the ray model for a non-telecentric AOL microscope and shows the experimentally observed distortion before correction. Experimental methods are described in section S5 and details on the correction of higher order field distortions after compensation for the non-telecentric distortions are given in S6.

## S1. PRECOMPENSATION OF THE REMOTE FOCUS DISTORTIONS

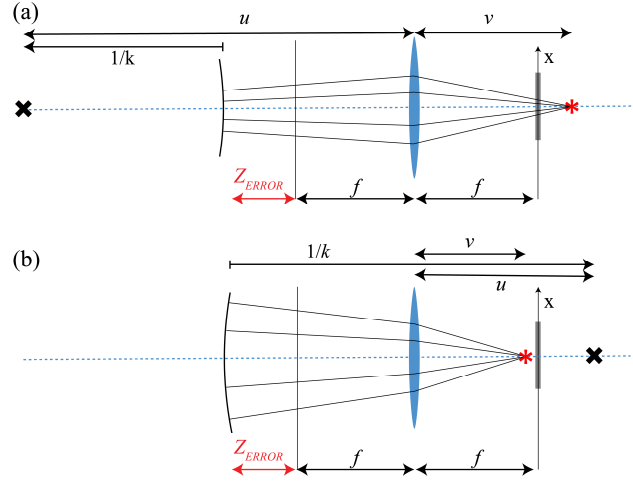

**Fig. S1. Remote focus for an axially misaligned remote focus unit.** Remote focus unit axially misaligned by an amount $Z_{ERROR}$ and objective lens of focal length $f$. **(a)** The diverging optical output of the remote focus unit results in a remote focus (*red* asterisk) below the natural focal plane of the objective (*grey* plane). The curved wavefront (*black* arc) converges at a point (*black* cross) a distance $1/\kappa$, where $\kappa$ is the curvature of the wavefront in m$^{-1}$. $u$ is the distance from the point of convergence to the objective lens and $v$ is the distance from the objective lens to the remote focus $z_{RF}$. **(b)** Same as (a) but for a converging optical output, resulting in a remote focus above the natural focal plane.

This section outlines the equations describing the distortion model and the correction method. In a remote focus system, the $z$-focus is controlled by the wavefront curvature $\kappa$ imparted to the beam by the remote focus unit. For $\kappa = 0$, the beam incident on the objective will be planar, resulting in a focus in the natural focal plane of the objective, at a distance one focal length from the objective. For a positive or negative $\kappa$, the objective input beam diverges [Fig. S1(a)] or converges [Fig. S1(b)], This is shown in Fig. S1 (a) and (b) for a single time point, where the diverging or converging wavefront results in a remote $z$-focus below or above the natural focal plane, respectively. The $z$-remote-focus in the objective field of view (FOV), $z_{RF}$ is the distance of the focus from the natural focal plane of the objective (i.e. when $\kappa = 0$) and may be found by applying the thin lens equation,

$$\frac{1}{f} = \frac{1}{u} + \frac{n}{v} \tag{S1}$$

where $f$ is the focal length of the paraxial objective, $u$ is the distance from the lens to the convergence point of the output of the remote focus unit, $v$ is the distance from the lens to the remote focus produced in the objective FOV and $n$ is the refractive index of the imaging medium. Applying equation S1 to the non-telecentric system shown in Fig. S1, where $Z_{ERROR}$ is the axial misalignment of the remote focus unit from its telecentric position, gives $u = (f + Z_{ERROR} + 1/\kappa)$ and $v = (nf – z_{RF})$ and solving for $z_{RF}$ gives:

$$z_{RF} = \frac{-nf^2 \kappa}{\kappa Z_{ERROR} + 1} \tag{S2}$$

From this it can be seen that in the case of an axially misaligned remote focus unit, $z_{RF}$ is no longer proportional to the wavefront curvature $\kappa$.

In Fig. 1 of the main text, $Z_d$ is the distance between the objective lens and the plane conjugate to the output of the remote focus unit, which can be found by applying equation S1:

$$Z_d = \frac{nf^2 + f Z_{ERROR}}{Z_{ERROR}} \tag{S3}$$

Moreover, the magnification, $M$ in the $xy$ plane is given by

$$M = \frac{Z_d - nf + z_{RF}}{Z_d - nf} \tag{S4}$$

and substituting the expression for $Z_d$ given by eq. S3, $M$ can then be expressed as

$$M = \frac{nf^2 + z_{RF} Z_{ERROR}}{nf^2} \tag{S5}$$

For the case of a remote focus unit misaligned in $(x,z)$ by an amount $(X_{ERROR}, Z_{ERROR})$, the position of the focus in the $x$-plane, $x_{RF}$, as measured from the optical axis depends on the projection of $X_{ERROR}$ along the tilted local $z$-axis and the varying lateral magnification, due to $Z_{ERROR}$ and is given by:

$$x_{RF} = X_{ERROR} \frac{z_{RF}}{nf} + \theta_X (f + Z_{ERROR})\left(\frac{Z_d - nf}{Z_d}\right) M \tag{S6}$$

where $\theta_x$ is the $x$ semi-scan angle of the beam by the remote focus unit. Substituting the expressions for $Z_d$ and $M$ given by equations S3 and S4, $x_{RF}$ may be rewritten as

$$x_{RF} = \frac{f(\theta_X - \kappa X_{ERROR})}{\kappa Z_{ERROR} + 1} \tag{S7}$$

Similarly, the $y$-focus in the objective FOV is given by,

$$y_{RF} = \frac{f(\theta_Y - \kappa Y_{ERROR})}{\kappa Z_{ERROR} + 1} \tag{S8}$$

where $\theta_y$ is the $y$ semi-scan angle.

Equations S2, S7 and S8 describe the distorted 3D focal position ($x_{RF}$, $y_{RF}$, $z_{RF}$) produced in the objective FOV, for a remote focus unit misaligned in $(x,y,z)$ by an amount ($X_{ERROR}$, $Y_{ERROR}$, $Z_{ERROR}$). By rearranging these equations for $\kappa$, $\theta_x$ and $\theta_y$, an expression can be found for the wavefront curvature and beam semi-scan angle that pre-compensates for the distortion, for a corrected $(x,y,z)$ focus at ($x_{CORR}$, $y_{CORR}$, $z_{CORR}$). These expressions, $\kappa_{COMP}$, $\theta_{x\ COMP}$ and $\theta_{y\ COMP}$, are integrated into the control software of the AOL and are given by,

$$\kappa_{COMP} = \frac{-z_{CORR}}{z_{CORR} Z_{ERROR} + nf^2} \tag{S9}$$

$$\theta_{X\ COMP} = \kappa_{COMP} X_{ERROR} + \frac{x_{CORR}(\kappa_{COMP} Z_{ERROR} + 1)}{f} \tag{S10}$$

$$\theta_{Y\ COMP} = \kappa_{COMP} Y_{ERROR} + \frac{y_{CORR}(\kappa_{COMP} Z_{ERROR} + 1)}{f} \tag{S11}$$

For remote focus systems containing a telecentric relay of magnification $M_R$ between the remote focus unit and objective, the expressions for $\kappa_{COMP}$, $\theta_{x\ COMP}$ and $\theta_{y\ COMP}$ are given by,

$$\kappa_{COMP} = \frac{-z_{CORR} M_R^2}{z_{CORR} M_R^2 Z_{ERROR} + nf^2} \tag{S12}$$

$$\theta_{X\ COMP} = \kappa_{COMP} X_{ERROR} + \frac{x_{CORR} M_R (\kappa_{COMP} Z_{ERROR} + 1)}{f} \tag{S13}$$

$$\theta_{Y\ COMP} = \kappa_{COMP} Y_{ERROR} + \frac{y_{CORR} M_R (\kappa_{COMP} Z_{ERROR} + 1)}{f} \tag{S14}$$

## S2. OPTICAL PATH OF THE REMOTE FOCUS ACOUSTO-OPTIC LENS 3D MICROSCOPE

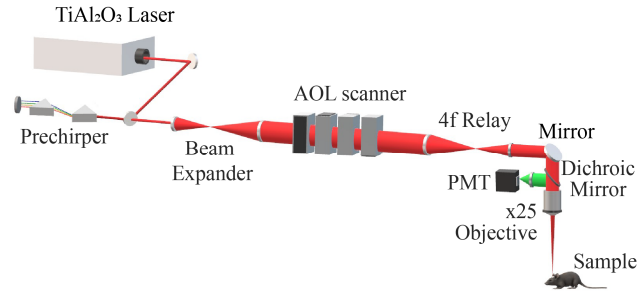

**Fig. S2. Optical path of the remote focus acousto-optic lens two-photon microscope.** A femtosecond pulsed laser generates a 2 W, 920 nm excitation beam (*red*). A prechirper precompensates temporal dispersion introduced by the acousto-optic lens (AOL) scanner. The AOL deflects and adds curvature to the optical beam, enabling remote focusing. A telecentric relay lens system relays the light beam to the back focal plane of the objective lens. Light emitted by two photon fluorescence excitation (*green*) at the remote focus within the sample is sent back through the objective lens to a photomultiplier tube (PMT) for data acquisition.

Fig. S2 shows the optical set-up of the acousto-optic lens (AOL) 3D two-photon microscope. A femtosecond pulsed laser (2 W at 920 nm; Chameleon Ultra II, Coherent Inc.) passes through a double pass prism-based pre-chirper before reaching the AOL 3D scanner. The custom-designed prism-based pre-chirper (APE GmbH, Berlin) introduces a group velocity dispersion of approx. 29,000 $fs^2$ to compensate for the temporal dispersion introduced by the AOL and the optical lenses in the system. The AOL consists of two orthogonally arranged pairs of $TeO_2$ AODs (Gooch and Housego) and comprises a compact geometry [1,2]. The AODs are interleaved with quarter-wave plates and polarizers to couple the beam into the subsequent AODs and to block the unwanted zero-order beam. The omission of inter-AOD telecentric relays that are included in earlier AOL designs [3,4], reduces the path length of the AOL from approx. 1.5 m to 20 cm. The AOL scanner deflects and adds curvature to the optical beam, which is subsequently relayed to the back focal plane of the objective lens by a relay lens system. The upright microscope consisted of an in-house optical arrangement mounted on top of a SliceScope (Scientifica, UK) with an telecentric relay, arranged to underfill a water-immersion objective (Leica HC FLUOTAR L 25X/0.95 W VISIR), giving an excitation NA of approx. 0.57. A two-channel detection system consisting of a dichroic mirror (575dcxr, Chroma Inc.) and emission filters (HQ 525/70m – 2P and HQ 630/100, - 2P), directed the red and green emission of the excited fluorophores onto two PMTs. Green fluorescence was detected with a GaAsP PMT (H7422, Hamamatsu, Japan) and red fluorescence with either a standard PMT (R9880U – 20, Hamamatsu, Japan) or a GaAsP PMT (H7422, Hamamatsu, Japan). The output signals from the PMTs were amplified using 200 MHz preamplifiers (Series DHPVA 100/200 MHz, FEMTO) and processed by a field programmable gate array (FPGA) based acquisition system. The acquisition system consisted of a high-speed ADC (800 MHz, dual channel, NI-5772) and an FPGA (NI FlexRIO, 7966R). A custom-designed FPGA-based control system generated the acoustic frequencies that drive the controlled operation of the AOL scanner. The FPGA AOL control system consisted of a Xilinx VC707 card and a Texas Instruments DAC card (DAC5672EVM). The commands to control the loading and execution of the acoustic drive frequencies were generated by a PC and encoded as RAW Ethernet packets before being transmitted to the AOL controller via a Gigabit Ethernet interface. The AOL control FPGA used an on-chip, direct digital synthesizer to generate the specified acoustic frequency chirps, which were executed upon receiving a start trigger from the data acquisition system. The synthesized digital waveforms were converted into analog signals by the DAC card and amplified by four RF amplifiers before being fed into the AODs. Further details on the design and operating principles of the compact AOL scanning remote focus two-photon microscope can be found in [1,2,5]

## S3. CALIBRATION STRATEGIES TO CORRECT FOR NON-TELECENTRIC MISALIGNMENT

The distortion calibration method we propose in this study used a layer of fluorescent beads to characterize the field distortion for each part of the field of view. During this calibration procedure, the image obtained at the natural focal plane was compared with an image obtained using remote focus. A mismatch between these two images caused by non-telecentricity in the optical path was then compensated. The values for $X_{ERROR}$, $Y_{ERROR}$ and $Z_{ERROR}$ were iteratively adjusted until a perfect registration between the predicted focus for a perfectly telecentric system (equivalent to the mechanical focus) and remote focus system was found. This calibration process was fast (a few minutes) and for the case of the AOL microscope, recalibration was only required if optical misalignments develop following movements of components in the optical path.

In practice (i.e. during an experiment), we did not need to re-evaluate the complete field distortion, but only verify that the introduction of additional optical elements (e.g. a coverslip) did not introduce additional distortions. In our experiments, ($X_{ERROR}$, $Y_{ERROR}$, $Z_{ERROR}$) could be easily measured in two steps, using a single fluorescent reference object in the sample (e.g. a small neuron or an injected fluorescent bead in a mouse brain) [Fig. S3]. Compensation of lateral misalignment must be done first. *C-z*-stack measurements of a reference point at the center of the FOV was used to calibrate and compensate lateral misalignment. Once lateral misalignment was corrected, moving the reference to the edge of the FOV could be used to calibrate magnification distortion and therefore compensate axial misalignment. For large axial misalignments, when the object may become out of focus, a manual adjustment can be used.

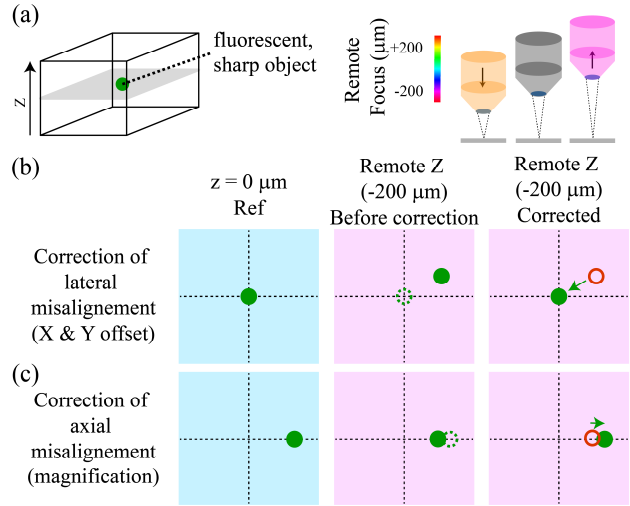

**Fig. S3. Schematic of lateral and axial misalignment estimation using only a single bright object. a)** *Left*: a small, bright object (*green*) used for the measurement is brought to the natural plane (*gray*). *Right:* the stage is moved mechanically, and the movement is compensated using remote focusing to keep the object in focus. **b)** Lateral misalignment is corrected first. *Left:* the object (*green*) is brought to the center of the FOV at *z* = 0. *Center:* The *z*-stage is then moved up by 200 µm and compensated by remote focusing -200 µm down (*pink* plane). If the object is shifted laterally compared to its expected position (*green dotted circle*), $X_{ERROR}$ and $Y_{ERROR}$ must be adjusted. *Right:* after successful correction, the object stays at the same *xy* location when remote focusing. **c)** Correction of axial misalignment. *Left:* the object is brought close to the edge of the FOV at *z*=0. *Center:* The *z*-stage is then moved up by 200 µm and compensated by remote focusing -200 µm down (*pink* plane). If the object is shifted laterally compared to its expected position (*green dotted circle*), this indicates a *z*-dependent magnification that can be corrected by adjusting $Z_{ERROR}$. *Right:* same as (b).

## S4. REMOTE FOCUS DISTORTIONS DUE TO NON-TELECENTRICITY

The main prediction of the paraxial distortion model is that any deviation of the output of the remote focus unit from its ideal telecentric position, results in a distorted FOV when focusing to any *z*-plane above or below the objective's natural focus. Fig. S4 and S5 shows the shape of the objective FOV (pink shaded volume), as predicted by the distortion model (i.e. eqs. S2, S7, S8) in MATLAB. In this example, the system consisted of an 8 mm focal length objective lens, a set of relay lenses of 0.6X magnification and a beam of wavefront curvature $\kappa$ ranging between ± 1 m$^{-1}$ and semi-scan angles $(\theta_x, \theta_y) = (5, 5)$ mrad. In Fig. S4 a lateral misalignment of the wavefront shaping of $(X_{ERROR}, Y_{ERROR}, Z_{ERROR})$ = (2,-2,0) mm, resulted in a skewed field of view (FOV) in the objective space. In Fig. S5, an axial misalignment of $(X_{ERROR}, Y_{ERROR}, Z_{ERROR})$ = (0,0,100) mm introduces a depth-dependent lateral magnification to the objective FOV, together with unequally spaced *z*-focal planes. A large $Z_{ERROR}$ was chosen for illustrative purposes. To further test the predictions of the remote focus distortion model, a separate paraxial ray model describing the AOL scanner and the subsequent focusing optics of the AOL microscope was developed in MATLAB. The acoustic drive equations used in the microscope control software were included in the ray model to simulate the remote focusing operation of the AOL for a 15 mm input beam of 920 nm wavelength. The modelled rays then pass through a relay of 0.6X magnification before being incident on a paraxial model of the Leica objective with a back-aperture diameter of 16 mm and focal length of 8 mm. The focus within the FOV below the objective was found by the least squares intersection point of the rays or by taking the point of maximum intensity of the point spread function given by the Fourier transform of the wavefront error in the iris plane of the objective. The former method was used in all figures showing the distorted bead trajectories predicted by the ray model and is sufficient for calculating the position of the distorted *xy* focus. The Fourier-based method was used in section S5 to predict the out-of-focus error resulting from 152 mm of $Z_{ERROR}$, when a more accurate *z*-focal position was required, however at the expense of increased computation time. Based on the AOL drive equations, the program could be set to simulate a volumetric remote focus raster scan and the shape of the FOV could be found by determining the extent of the set of foci produced during the volumetric scan. The position of the AOL unit in the ray model can be misaligned, resulting in a distorted FOV. In Fig. S4 and S5, the blue rays of the ray model are overlaid with the FOV predicted by the distortion model and are shown forming a focus at (*x,y,z*) = (0,0,0) µm. The black asterisks indicate the foci produced at the center of the FOV for increments of 0.1 m$^{-1}$ in the wavefront curvature $\kappa$. The foci of the ray model at the 3D extreme of the simulated raster scan, shown by the red asterisks in Fig. S4 and S5, coincide with the distorted FOV predicted by the distortion model.

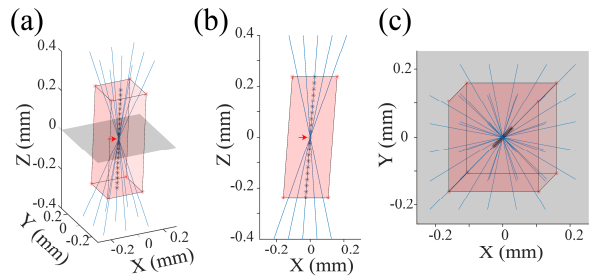

**Fig. S4. Predicted effect of lateral misalignment on the 3D FOV.** Ray model and distortion model for AOL misalignment ($X_{ERROR}, Y_{ERROR}, Z_{ERROR}$) = (2,-2,0) mm. **a)** 3D view of the objective FOV predicted by the distortion model (*pink*), overlaid with the rays of the ray model (*blue*) focusing at (x,y,z) = (0,0,0) μm (*red arrow*). Black asterisks indicate the center of the FOV at various remote focus planes. *Grey* plane perpendicular to the FOV indicates the natural focal plane of the objective (z = 0). Red asterisks indicate the foci produced by the ray model at the corners of the 3D FOV and match the corners of the pink FOV produced by the distortion model. **b)** same as (a) for the side view. **c)** same as (a) for the top view. Note the skew arising from the lateral misalignment of the AOL.

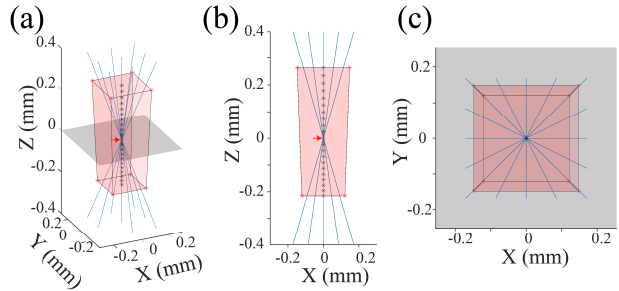

**Fig. S5. Predicted effect of axial misalignment on the 3D FOV.** Ray model and distortion model for AOL misalignment ($X_{ERROR}, Y_{ERROR}, Z_{ERROR}$) = (0,0,100) mm. **a)** 3D view of the objective FOV predicted by the distortion model (*pink*), overlaid with the rays of the ray model (*blue*) focusing at (x,y,z) = (0,0,0) μm (*red arrow*). *Black* asterisks indicate the center of the FOV at various focus planes. *Grey* plane perpendicular to the FOV indicates the natural focal plane of the objective (z = 0). Red asterisks indicate the foci produced by the ray model at the corners of the 3D FOV and match the corners of the *pink* FOV produced by the distortion model. **b)** same as (a) for the side view. **c)** same as (a) for the top view. Note varying z-plane spacing and varying xy plane magnification arising from the axial AOL misalignment.

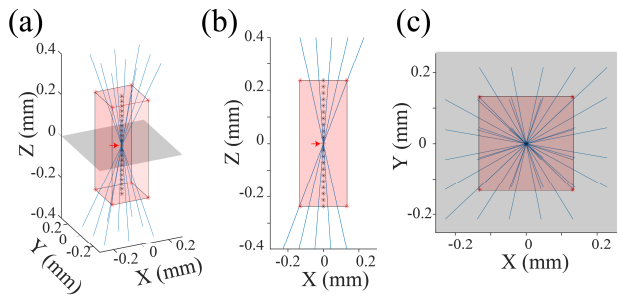

**Fig. S6. Correction of predicted axial and lateral misalignment using distortion precompensation.** Ray model and distortion model for AOL misalignment ($X_{ERROR}, Y_{ERROR}, Z_{ERROR}$) = (2,-2,100) mm with distortion precompensation. **a)** 3D view of the objective FOV predicted by the distortion model (*pink*, using $X_{ERROR}, Y_{ERROR}, Z_{ERROR}$ = (0,0,0) mm), overlaid with the rays of the precompensated ray model (*blue*) focusing at (x,y,z) = (0,0,0) μm (*red* arrow). Black asterisks indicate the center of the FOV at various remote foci. Gray plane perpendicular to the FOV indicates the natural focal plane of the objective (z = 0). *Red* asterisks indicate the foci produced by the ray model at the corners of the 3D FOV and match the corners of the pink FOV produced by the distortion model. **b)** same as (a) for the side view. **c)** same as (a) for the top view. Note that distortions observed in figures S3 and S4 are corrected but that the central ray remains skewed.

The ray model was also used to simulate the operation of the distortion correction scheme by modelling the adjustments to the acoustic drive frequencies required for precompensation. Fig. S6 shows the ray foci produced by the ray model within the objective FOV with distortion precompensation for an AOL misalignment of ($X_{ERROR}, Y_{ERROR}, Z_{ERROR}$) = (2, -2,100) mm. The pink shaded FOV is the FOV predicted by the distortion model for a telecentric AOL (i.e. ($X_{ERROR}, Y_{ERROR}, Z_{ERROR}$) = (0,0,0) mm). Fig. S6 demonstrates that the distortion pre-compensation scheme, which is based on the inverse paraxial distortion model, can be used to correct for remote focus distortions in any optical set-up described by a varying wavefront curvature, a paraxial objective of any focal length and, optionally, a set of relay lenses. The ray model can also be used to estimate the real misalignment of the AOL in the microscope by matching the experimentally measured and the theoretically predicted distortion. This feature can be applied to any remote focus microscope provided a *C-z*-stack can be obtained. The estimate of the non-telecentric misalignments in the system given by the ray model can be used to physically align the remote focus components to compensate for the observed distortions. Fig. S7 shows the experimentally measured distortions due to the lateral (S4) and axial (S5) misalignments of the AOL. From this, the AOL misalignment inherent to our remote focus microscope, ($X_{ERROR\_AOL}$, $Y_{ERROR\_AOL}$, $Z_{ERROR\_AOL}$), was estimated to equal (1.4, 0.4, 69) mm.

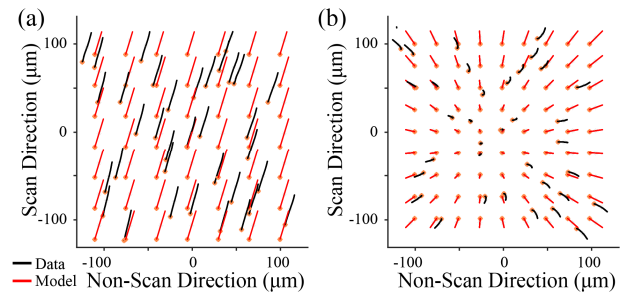

**Fig. S7. Experimentally measured and predicted lateral and axial misalignment. a)** Measured bead position trajectories (*black*) and bead positions predicted by the ray model (*red*) due to inherent lateral misalignment of our AOL microscope. *Orange* and *gray* asterisks indicate values for 200 μm remote focusing and natural plane imaging, respectively. Axial precompensation was applied to isolate the effect of lateral misalignment. **b)** same as (a) showing inherent axial misalignment of our AOL microscope. Lateral precompensation was applied to isolate the effect of axial misalignment.

## S5. TESTING THE DISTORTION CORRECTION SCHEME

A schematic of the optical set-up used to test the performance of the distortion correction scheme for effective lateral misalignments of the AOL, is given in Fig. S8. A pair of Risley prisms were placed within the relay preceding the objective lens, to mimic lateral displacements of the AOL scanner. The amount of beam displacement introduced by the Risley prisms could be calculated as a function of their rotational position. *C-z*-stacks were then collected with and without the correction compensation scheme enabled. The lateral beam displacement by the Risley prisms lead to a skewed objective FOV, which can be measured by taking the inverse tangent of the slope of the bead trajectories across a *C-z*-stack. The effectiveness of the distortion correction scheme was determined by comparing the experimentally measured skew in the FOV with and without the correction system enabled. As described in section 3 of the main text, the distortion correction scheme is shown to be effective for lateral misalignments of the remote focus unit of up to 7 mm.

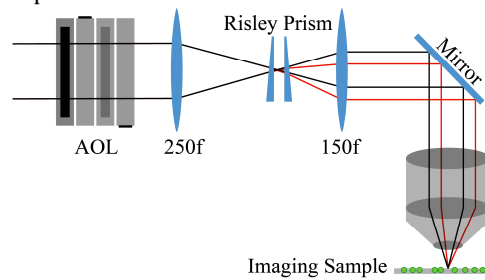

**Fig. S8. Schematic of optical setup for introducing different degrees of lateral AOL misalignment.** Telecentric condition, where the light beam is centered in the absence of Risley prisms (*black lines*). Light beam when a pair of Risley prisms are placed within the telecentric relay preceding the objective, mimicking a lateral misalignment of the remote focusing device. This causes a lateral displacement of the beam (*red*) on the back aperture of the objective and a skew of the imaging FOV.

To test the effectiveness of the distortion correction scheme for axial misalignments of the remote focus unit, the lenses within relay subsequent to the AOL scanner were modified in order to change the magnification of the optical path and in this way introduced an effective change of the axial misalignment of the AOL unit. By placing a lens of -750 mm and +750 mm focal length directly after the 250 mm focal length lens in the relay, the effective focal length of the first relay lens was changed to 375 mm and 188 mm, respectively. In addition to changing the magnification of the optical system subsequent to the AOL, this introduced an extra divergence and convergence to the input incident on the back aperture of the objective, thus shifting the absolute focal plane. Fig. S9 shows the effect of changing the effective focal length of the first relay lens in the relay. By comparing the experimentally observed distortion to that predicted by the ray model, it is estimated that the addition of the -750 mm *f* and +750 mm *f* lens introduced an effective additional axial misalignment of the AOL equal to +83 mm and -60 mm, respectively. Fig. S10(a,c) shows the experimentally observed remote focus distortion introduced by the -750 mm *f* and +750 mm *f* lens, respectively, with correction for the approx. 69 mm inherent axial misalignment in the system. Fig. S10(b,d) shows the effectiveness of the distortion precompensation scheme in correcting for these effective misalignments. Fig. S10(e,f) shows the comparison between the experimentally observed distortion and that predicted by the ray model, which was used to estimate the amount of additional effective axial misalignment introduced by the addition of the -750 mm $f$ and +750 mm $f$ lens.

To confirm that the distortion pre-compensation scheme, which corrects for the magnification distortion introduced by axial misalignments of the AOL also corrects for the non-constant $z$-spacing, the $z$ focal plane position was measured before and after correction by the distortion precompensation scheme [Fig. S11]. Using a sample of 0.2 µm beads suspended in polymer, the AOL was used to remote focus at $z$ = [-200, 0, 200] µm as described previously.

The objective was then mechanically displaced between 30 µm above and 30 µm below each of the remote focus positions, in 1 µm increments, to find the true $z$ focal plane position with and without the distortion precompensation scheme enabled. The out-of-focus error arising from the pre-existing axial AOL misalignment of 69 mm was measured to be -9.5 µm and 13.5 µm for a remote focus of -200 µm and 200 µm, respectively, which was consistent with the error predicted by the Fourier-based ray model, -10.0 µm and 13.9 µm. With the distortion precompensation scheme enabled, the out-of-focus error was reduced to -1.5 µm and 1.5 µm, respectively. For an axial AOL misalignment of 152 mm (the case shown in Fig. S9 (a,c,e)), the out-of-focus error predicted by the Fourier-based ray model for a remote focus of [-200, 200] µm was found to be [-23.5, 32.5] µm.

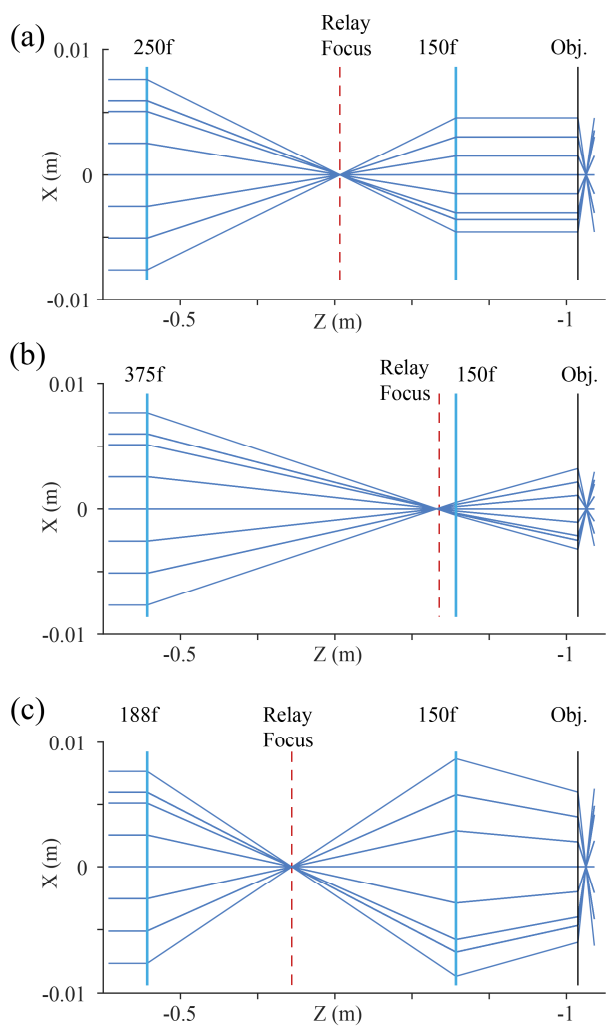

**Fig. S9. Ray model diagrams of the modified relay for a collimated AOL output. a)** The regular relay (with 2 relay lenses, 250 $f$ and 150 $f$, giving 0.6X magnification) present in our AOL microscope. **b)** same as (a), adding a -750 mm focal length lens after the first relay, akin to having a single 375 $f$ lens. This introduces an additional beam convergence (magnification = 0.4X). **c)** same as (a), adding a +750 mm focal length lens after the first relay, akin to having a single 188 $f$ lens. This introduces an additional beam divergence (magnification = 0.8X).

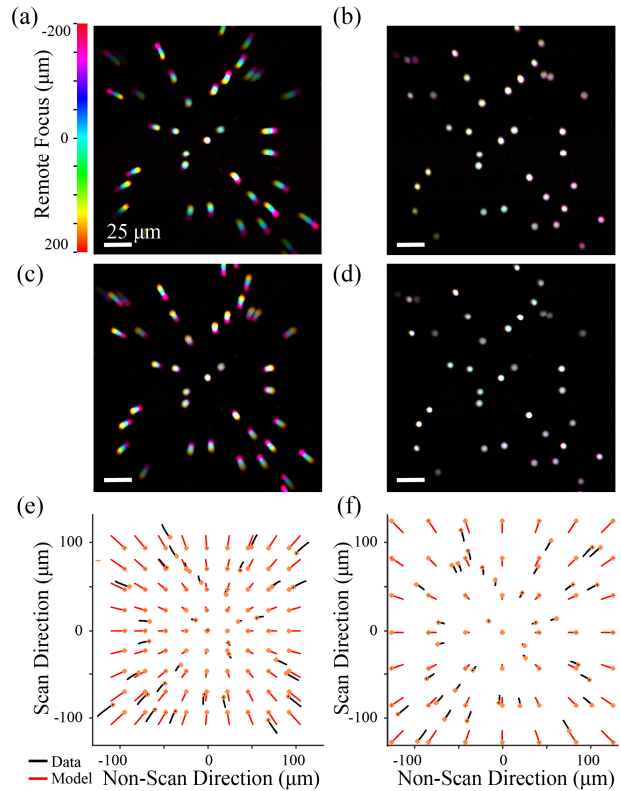

**Fig. S10. Experimental test of distortion correction scheme for axial AOL misalignments: a)** Mean intensity projection of imaged beads at various remote foci (± 200 μm in 5 μm increments, color indicates z-plane location). Colored smearing indicates depth-dependent deviation of the bead location relative to the natural plane image. The +83 mm of additional effective axial misalignment (see Fig. S9b) causes z-dependent magnification of the FOV. Lateral misalignment was pre-compensated. **b)** Same as (a) with axial distortion precompensation enabled **c)** Same as (a), with introduction of -60 mm of additional axial AOL misalignment (See Fig S9c). **d)** same as (c) with distortion precompensation enabled. **e)** Bead position trajectories tracked using ImageJ TrackMate plugin (*black*) and bead positions predicted by the ray model (*red*) for (a), with $Z_{ERROR}$ = +83 mm. *Orange* and *gray* asterisks indicate values for 200 μm remote focusing and natural plane imaging, respectively. **f)** same as (e), with $Z_{ERROR}$ = -60 mm.

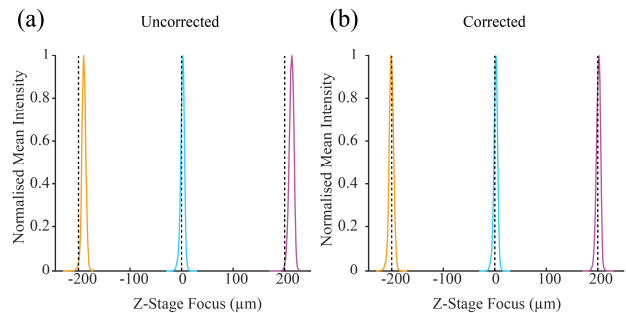

**Fig. S11. Experimental correction of z-focus error due to axial misalignment. a)** Measure of the *z*-focus error in our AOL system when remote focusing at +200 and -200 μm, using 0.2 μm beads. Out-of-focus error was -9.5 and +13.5 μm, respectively. **b)** Correction for axial misalignment (cf. Fig. S10) also corrected non-constant spacing of the *z* planes. Residual out-of-focus error is -1.5 and +1.5 μm.

## S6. APPLICATIONS OF THE DISTORTION CORRECTION SCHEME

The high precision of the distortion precompensation scheme allowed us to correct for residual second order distortions that could only be observed after distortion precompensation for non-telecentricity in the AOL microscope. This second order distortion was found to vary in a non-uniform manner across the FOV and to increase with longer line scan times. Fig. S12(a) shows the distorted trajectories of 5 μm fluorescent beads across a 270 x 270 x 400 μm FOV, measured with a 200 ns pixel dwell time for a 512x512 pixel scan.

The origin of this distortion was investigated using the ray model and was found to arise from a non-parallel input beam to the AOL. Figs. S12(a-c) show the match between the experimental trajectories and those predicted by the ray model for a modelled diverging AOL input of -0.1 m$^{-1}$ curvature. This prediction allowed us to realign the optics preceding the AOL, leading to a sub

0.5 μm mean positional error over a ± 200 μm remote focus range, independent of pixel dwell time, where the distortion error was calculated as the maximum absolute displacement of the bead trajectory from its position at $z = 0$ in the outer regions of the FOV [S12(e)]. Figure S12 (d) shows the trajectories of the beads in (a), after realignment of the AOL input beam. Fig. S12(f) gives the precision achieved over a range of dwell times and remote focus ranges. The slight rise in error at the higher dwell times for the ± 200μm remote focus range was caused by an AOL aperturing effect that arises when the AODs within the AOL reach their frequency limits for efficient diffraction. This affected the outer regions of the FOV for longer dwell times at increasing $z$ focus. For this reason, the mean error against dwell time is also given for a ±100 μm $z$ range, showing that realignment of the AOL input beam resolved the dwell time dependence of the residual distortion. This correction highlights that the high precision of the distortion precompensation scheme allows smaller, higher order distortions to be identified and corrected. The correction for beam divergence into the AOL was only possible after precompensation for the distortion arising from system non-telecentricity.

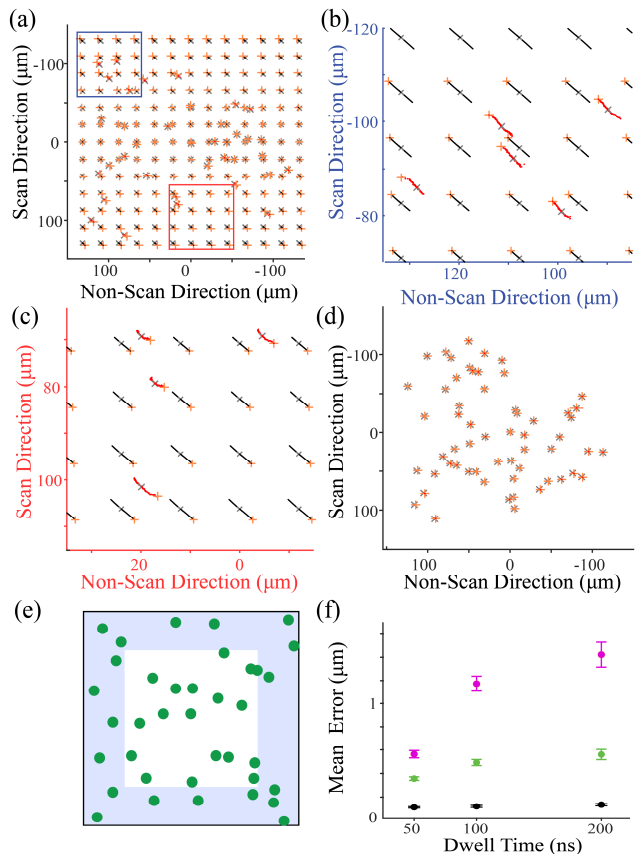

**Fig. S12. Properties of residual distortion in AOL FOV and its correction. a)** Trajectories of 5 μm fluorescent beads (*red*) over a ±200 μm remote focus range, after distortion precompensation for AOL misalignment. Taken at a 200 ns pixel dwell time for a 512 x 512 pixel FOV. Distortion predicted by the ray model (*black*), for an AOL input beam divergence of -0.1 m$^{-1}$. *Orange* and *gray* crosses indicate values for 200 μm remote focusing and natural plane imaging, respectively, for both the experimental and theoretical data. **b)** Zoom of the region in (a) marked by the *blue* outline with experimental (*red*) and modelled (*black*) trajectories. **c)** Zoom of the region in (a) marked by the *red* outline. **d)** Trajectories of the beads in (a) after realignment of the AOL input beam with zero divergence. **e)** Region of FOV where the field distortion is greatest (blue shading). Beads outside of this region were excluded from the mean distortion error calculation. **f)** Mean positional error of 5 μm bead trajectories in a 270 x 270 μm FOV against dwell time, before (*purple*) and after (*green*) realignment of the AOL input beam for a ±200 μm remote focus range and for a ±100 μm remote focus range (*black*). Error bars show standard error.